\documentclass[useAMS,usenatbib]{mn2e}
\usepackage{amsmath,fleqn,graphicx,amssymb,xcolor}
\usepackage{multirow}
\renewcommand{\[}{\begin{equation}}
\renewcommand{\]}{\end{equation}}
\def\p{\partial}\def\i{{\rm i}}

\def\rd{}

\let\boldgrk=\gkvecten
\let\boldgrksc=\gkvecseven

\def\gkthing#1{{\mathchoice%
	{\hbox{{\boldgrk\char#1}}}
	{\hbox{{\boldgrk\char#1}}}
	{\hbox{{\boldgrksc\char#1}}}
	{\hbox{{\boldgrksc\char#1}}}}}

\def\vtheta{\gkthing{18}}

\def\vxi{\gkthing{24}}

{\newif\ifnotend
\notendtrue
\def\veclist{ABCDEFGHIJKLMNOPQRSTUVWXYZabcdefghijklmnopqrstuvwxyz.}
\def\top#1#2.{#1}
\def\tail#1#2.{#2.}
\loop\expandafter\xdef\csname v\expandafter\top\veclist\endcsname%
{{\noexpand\bf\expandafter\top\veclist}}
\edef\veclist{\expandafter\tail\veclist}
\if\veclist.\notendfalse\fi\ifnotend\repeat}
{\newif\ifnotend
\notendtrue
\def\callist{ABCDEFGHIJKLMNOPQRSTUVWXYZ.}
\def\top#1#2.{#1}
\def\tail#1#2.{#2.}
\loop\expandafter\xdef\csname c\expandafter\top\callist\endcsname%
{{\noexpand\cal\expandafter\top\callist}}
\edef\callist{\expandafter\tail\callist}
\if\callist.\notendfalse\fi\ifnotend\repeat}

\def\ex#1{\left<#1\right>}

\def\ket#1{|#1\rangle}
\def\la{\langle}

\def\d{{\rm d}}

\def\kB{k_{\rm B}}

\def\bolOm{\mbox{\boldmath$\Omega$}}
\def\vOmega{\bolOm}

\def\e{\mathrm{e}}

\def\fracj#1#2{{\textstyle{#1\over#2}}}

\def\eqrf#1{(\ref{#1})}

\title[Probabilistic distribution functions]
{Probabilistic distribution functions}

\author[Jun Yan Lau \& James Binney]{
  Jun Yan Lau$^{1,2}$\thanks{E-mail: jun.lau.20@ucl.ac.uk} \& James Binney$^2$\thanks{E-mail:
  binney@physics.ox.ac.uk}\\   
  $^1$UCL Mullard Space Sciences Laboratory, Holmbury St Mary, Surrey RH5
  6NT\\
  $^2$Rudolf Peierls Centre for Theoretical Physics, Clarendon Laboratory,
  Parks Road, Oxford, OX1 3PU, UK
}

\begin{document}
\maketitle

\begin{abstract}
Observed clusters  should be modelled by considering the distribution
function to be a random variable that quantifies the degree of excitation of
the system's normal modes. A system of canonical coordinates for the space of
DFs is identified so DFs can be weighted in a consistent way.\end{abstract}


\section{Introduction} \label{sec:intro}

Galaxies and star clusters are in approximate states of equilibrium and have
for decades been fitted to models in which the distribution function (DF)
$f(\vx,\vv)$ of their constituent particles (stars, dark-matter particles)
are steady-state solutions of the collisionless Boltzmann equation (CBE).
The advent of massive simulations of galaxy formation \citep[e.g.][]{Laporte2019} and
detailed data from the Gaia mission \citep{GaiaDR2general} and large integral
field units such as MUSE \citep[e.g.][]{MUSE2021} have stimulated interest in non-equilibrium
features of galaxies, especially the Milky Way \citep[e.g.][]{AntojaSpiral}. 

It is likely that equilibrium models will not suffice to explain the
exquisite data that Gaia is delivering because the data, effectively taken at
one instant of a cluster's life, capture a specific fluctuation. Hence we now
need models that embrace fluctuations. Clearly this can only be done in a
statistical sense: the requirement is to predict which deviations from a
mean-field model are likely. That is, we need to assign probabilities to
distribution functions \citep{Ma06}, and the theory will be tested by comparing with
observations its predictions for populations of clusters  or galaxies.

The dynamics of a crystal are often best studied by going to the continuum
limit in which quantities like the displacement $\vxi$ of atoms from equilibrium
positions and the orientations $\vs$ of spins are defined at all $\vx$. Then the state
of the system is defined by fields such as $\vxi(\vx)$ or $\vs(\vx)$.
Similarly, in stellar dynamics we want to characterise the disturbance of a
cluster by the field $f(\vw)$, where $\vw=(\vx,\vv)$ is a location in phase
space. Then we want to compute expectation values
of observables $\cO$ by taking expectation values over all possible fields $f$:
doing so we are computing a double expectation: first
$\ex{\cO}_{f}=\int\d^6\vw\,f(\vw)\cO(\vw)$ and then an average of these averages
weighted by the probability of each DF $f$. 

An example of an important observable to predict in this way is the variance of a suitable
dipole moment to quantify the tendency for the densest part of a
cluster to be offset from the cluster's barycentre. In this case one would
take $\cO$ to be the product of a function $R(r)$ that has opposite signs at
$r=0$ and $r\to\infty$ and an $\ell=1$ spherical harmonic:
\[
\cO(\vw)=R(r)Y_1^m(\theta,\phi).
\]
In the case of an isolated, {\rd non-rotating} cluster $\ex{\cO}_f$ would then be a random
variable with vanishing mean and a dispersion that would be independent of
the azimuthal quantum number $m$. The variance of $\ex{\cO}_f$ would be a
prediction that could be tested from observations of a sufficient number of
real or simulated clusters \citep[e.g.][]{LauBinney2019,Heggie2020}.

This programme requires a rule for assigning a priori probabilities to
possible fields $f(\vw)$. In classical statistical mechanics the analogous
rule is inferred by noting that the a priori probability of some range of
phase-space locations $\vw$ must be independent of time as the system evolves
undisturbed, and Liouville's theorem ensures that this condition is satisfied
if a priori probability is proportional to the volume element
$\d^6\vw=\d^3\vq\d^3\vp$ defined by a system of canonical coordinates
$(\vq,\vp)$ -- here
it's important that (1) any system of canonical coordinates assigns
probability in the same way, and (2) the assignment is time-independent
because Hamilton's equations effect a series of canonical transformations.

{\rd Guided by this analogy, we seek a system of canonical coordinates for the
space of possible DFs.}
Canonical coordinates are defined to be those in which Poisson brackets
$[f,g]$ have the canonical form
\[
[f,g]=\sum_i\bigg({\p f\over\p\vq}\cdot{\p g\over\p\vp}-
{\p f\over\p\vp}\cdot{\p g\over\p\vq}\bigg).
\]
Poisson brackets impose a symplectic structure on phase space and canonical
coordinates are privileged coordinates with respect to this structure in the
same way that Cartesian coordinates are privileged with respect to a
Euclidean metric \citep[e.g.][]{Arnold}.
So the key to defining a field theory for stellar dynamics is imposing a
symplectic structure on the space of all possible DFs. That is, we must
define an antisymmetric bilinear form $\{,\}$ that yields a functional on the
space of DFs when the two slots are filled by two DFs. The standard
phase-space Poisson bracket is not a candidate for this job because
$[f,g]$ is a function on phase space rather than a functional. In Section
\ref{sec:symp} we identify a symplectic structure for the space of DFs.
This structure then allows us to identify in Section \ref{sec:qp} canonical coordinates for the space
of DFs.

LB21 have recently argued that the fluctuations of a stellar
system are best tackled in terms of the system's van Kampen modes.  They
showed that a system's excitation energy is the sum of the energies invested
in each of its van Kampen modes. The energy of a growing or decaying mode is
identically zero and in Section \ref{sec:E} we show that when the
energy of a stable mode is expressed in terms of canonical coordinates, it
takes the form of a sum of harmonic-oscillator Hamiltonians. This result
suggests that it may be possible to apply to clusters the methods of
classical equilibrium statistical mechanics.

\section{Mathematical background and notation}

Here we introduce two vital concepts and define our notation.

\subsection{Angle-action variables} Actions $J_i$ are constants of motion
that can serve as canonical phase-space coordinates. 
Their conjugate variables, the angles $\theta_i$, increase linearly in
time, so $\vtheta(t)=\vtheta(0)+\vOmega t$. A particles' Hamiltonian
$H(\vx,\vv)$ is a function $H(\vJ)$ of the actions only and the frequencies
$\Omega_i$ that control the rates of increase of the angles are given by
$\vOmega=\p H/\p\vJ$. Since angle-action variables are canonical coordinates,
the element of phase-space volume
$\d^6\vw=\d^3\vx\,\d^3\vv=\d^3\vtheta\,\d^3\vJ$.
Functions on phase space can be expressed as Fourier series:
\[\label{eq:defsFT}
h(\vw)=\sum_\vk h_\vk(\vJ)\e^{\i\vk\cdot\vtheta}\ ;\ 
h_\vk(\vJ)=\int{\d^3\vtheta\over(2\pi)^3}\e^{-\i\vk\cdot\vtheta}h(\vw).
\]
Note that for real $h$, $h_{-\vk}=h_\vk^*$.

\subsection{van Kampen modes}

The CBE is time-reversal symmetric, so group-theoretic arguments suggest that
it should have a complete set of solutions with time dependence $\e^{\i\omega
t}$, with $\omega$ possibly complex. These solutions are called van Kampen
modes. \citet[hereafter LB21]{LauBinney2021a} show that these
solutions are the eigenfunctions of a Hermitian operator on phase space $K$,
with eigenvalue $\omega^2$. Since the eigenvalues of a Hermitian operator are
all real, $\omega$ is either real or pure imaginary. Time-reversal symmetry
implies that if there is a van Kampen mode with frequency $\omega=\i y$ with
$y$ real, then there is another mode with frequency $\omega^*=-\i y$. Hence
modes are  either pure oscillatory or exponentially growing/decaying in
pairs. The real frequencies form a continuum, while the imaginary frequencies
are isolated.

\section{A symplectic structure for the space of possible DFs}
\label{sec:symp}

In this section we identify the symplectic structure for the space of DFs.
Coordinates for the space of DFs are numbers that characterise a DF in the
same way that quantum amplitudes $a_i\equiv \la E_i\ket\psi$ are numbers that
characterise the state $\ket\psi$ of a quantum system. Suitable numbers can
be extracted by taking an inner product with a function $a_i(\vw)$:
\[
\hat a_i[f]\equiv\big(a_i,f\big)\equiv\int\d^6\vw\,a_i^*(\vw)f(\vw).
\]
Here a hat implies a functional on the space of DFs, and the square brackets
imply that the argument is a function on phase space rather than a number.
One possible choice of function that defines $\hat a_i$ is
$a_{\vw'}(\vw)=\delta(\vw-\vw')$. Then $\hat a_{\vw'}[f]=f(\vw')$ is just the
value of the DF at the location $\vw'$ associated with this coordinate.
Another widely employed choice is
$a_{\vk\vJ'}=\delta(\vJ'-\vJ)\e^{-\i\vk\cdot\vtheta}/(2\pi)^3$, which yields
$\hat a_{\vk\vJ'}[f]=f_\vk(\vJ')$.  We show below that {\rd via
equation \eqrf{eq:def_pq}} the $f_\vk$ yield canonical coordinates for the space of DFs.

We are interested in how coordinates vary as we move about the space of DFs,
so we need the functional derivative
\[
{\delta\hat a_i\over\delta
f(\vw')}\equiv\int\d^6\vw\,a_i^*(\vw)\delta(\vw-\vw')=a_i^*(\vw').
\]
Note that taking a functional derivative of a coordinate $\hat a_i$ (a
functional) we
obtain a function on phase space.

Inspired by \cite{Morrison1980}, we construct a symplectic operator
such that CBE can be written in the form of Hamilton's equations, namely
\[\label{eq:newCBE}
{\p \hat f\over\p t}=\{\hat f,\hat H_{\rm s}\},
\]
where $\hat f$ is a functional such as $\hat a_i$ that characterises the
system's 
DF and\footnote{$\Phi$ appears with a factor half in the Hamiltonian because it is the
potential self-consistently generated by the DF rather than an externally
imposed potential.}
\[
\hat H_{\rm s}[f]\equiv\fracj12\int\d^6\vw\,f(\vw)\big(v^2+\Phi(\vw)\big),
\]
with
\[
\Phi(\vw)=-GM\int\d^6\vw'\,{f(\vw')\over|\vx-\vx'|}.
\]
In equation \eqrf{eq:newCBE} both sides are functionals and  the Poisson bracket on the right is
defined by
\[\label{eq:defsPB}
\{\hat U,\hat V\}[f]\equiv\int\d^6\vw\,f(\vw)\bigg[{\delta\hat U\over\delta
f(\vw)},{\delta\hat V\over\delta f(\vw)}\bigg],
\]
where the square brackets in the integral denote a conventional Poisson bracket.

\subsection{Equivalence of equation \eqrf{eq:newCBE} and the CBE}

The functional derivative of $\hat H_{\rm s}$ is
\begin{align}\label{eq:fderivH}
{\delta\hat H_{\rm s}\over\delta f(\vw')}&=\fracj12\left(
v^{\prime2}+\Phi(\vx')\right)+\fracj12\int\d^6\vw\,f(\vw){\delta\Phi(\vw)\over\delta
f(\vw')}\cr
&=\fracj12\left(
v^{\prime2}+\Phi(\vx')\right)-\fracj12GM\int\d^6\vw\,{f(\vw)\over|\vx-\vx'|}\cr
&=\fracj12
v^{\prime2}+\Phi(\vx'),
\end{align}
which is the Hamiltonian of a single star.
When we substitute the Poisson bracket's definition \eqrf{eq:defsPB} into
equation
\eqrf{eq:newCBE} and use our
expression for the functional derivative of $\hat H_{\rm s}$, we find 
\[
{\p \hat f\over\p t}[f]=
\int\d^6\vw\,f(\vw)\bigg[{\delta\hat f\over\delta
f(\vw)},H_{\rm s}(\vw)\bigg],
\]
This equation holds for any choice of functional (coordinate) $\hat f$ on DF
space. If we set $\hat f$ equal to the functional $\hat a_{\vw'}$ with
$a_{\vw'}(\vw)=\delta(\vw-\vw')$ explored above, then the number $\hat f[f]$
becomes $f(\vw')$ and the functional derivative $\delta\hat f/\delta
f(\vw)$ becomes $\delta(\vw-\vw')$, so we have
\[\label{eq:nearly}
{\p f(\vw')\over\p t}=\int\d^6\vw\,f(\vw)[\delta(\vw-\vw'),H(\vw)].
\]
 For any three functions we have that
$\int\d^6\vw\,f[g,h]=\int\d^6\vw\,g[h,f]$ (provided one or more vanishes at
infinity), so equation \eqrf{eq:nearly} yields the  conventional CBE
\[\label{eq:oldCBE}
 {\p f(\vw')\over\p t}=[H(\vw'),f(\vw')],
\]
The opposing signs in these two forms of the CBE \eqrf{eq:newCBE} and
\eqrf{eq:oldCBE} reflect the difference between Hamilton's equation $\dot
q=[q,H]$ and the conventional CBE $0=\dot f=\p_tf+[f,H]\Rightarrow\p_t
f=-[f,H]$. That is, equation \eqrf{eq:newCBE} is the equation of motion of
the DF's coordinates, while equation \eqrf{eq:oldCBE} gives the rate of
change of the value of $f$ at fixed $\vw$.

Above we derived the conventional CBE \eqrf{eq:oldCBE} from equation
\eqrf{eq:newCBE}, but reversing the chain or arguments one can also show that
equation \eqrf{eq:newCBE} follows from the conventional CBE; the statements
$\p_t \hat f=\{\hat f,\hat H_{\rm s}\}$ and $\p_tf=-[f,H]$ are equivalent.

\subsection{Application to equilibrium systems}\label{sec:qp}

We now examine the structure of the functional Poisson bracket when evaluated
on a DF that differs from that of an equilibrium model only by virtue of a
fluctuation. That is, we consider DFs of the form
\[
f(\vw)=f_0(\vJ)+f_1(\vtheta,\vJ).
\]
In this case the Poisson bracket
{\rd\eqrf{eq:defsPB}} can be written
\begin{align}\label{eq:defsLPB}
&\{\hat U,\hat V\}[f]=\int\d^6\vw\,{\delta\hat U\over\delta
f(\vw)}\bigg[{\delta\hat V\over\delta f(\vw)},f\bigg]\cr
&=\int\!\d^6\vw\,{\delta\hat U\over\delta
f(\vw)}\,\bigg({\p\over\p\vtheta}{\delta\hat V\over\delta f(\vw)}\cdot{\p
f\over\p\vJ}-{\p\over\p\vJ}{\delta\hat V\over\delta f(\vw)}\cdot{\p
f\over\p\vtheta}\bigg).
\end{align}
The second term in the round bracket is smaller than the first by
O($f_1/f_0$), so to leading order we have
\[\label{eq:lo}
\{\hat U,\hat V\}[f]=\int\d^6\vw\,{\delta\hat U\over\delta
f(\vw)}\,{\p
f_0\over\p\vJ}\cdot{\p\over\p\vtheta}{\delta\hat V\over\delta f(\vw)}
\]
If we define
\begin{align}
x_\vk&=\Re(f_\vk)=\int{\d^3\vtheta\over(2\pi)^3}f\cos(\vk\cdot\vtheta)\cr
y_\vk&=\Im(f_\vk)=-\int{\d^3\vtheta\over(2\pi)^3}f\sin(\vk\cdot\vtheta),
\end{align}
then we can restrict sums over $\vk$ to a half-plane, such as $k_1\ge0$ and
\[
{\delta x_\vk(\vJ)\over\delta f(\vw)}={\cos{\vk\cdot\vtheta}\over(2\pi)^3}
\hbox{ and }
{\delta y_\vk(\vJ)\over\delta f(\vw)}=-{\sin{\vk\cdot\vtheta}\over(2\pi)^3}.
\]
Now applying the chain rule to \eqrf{eq:lo} yields
\begin{align}\label{eq:PBUVlin}
\{\hat U,&\hat V\}[f]={1\over(2\pi)^6}\int\d^6\vw\cr
&\sum_\vk\Big(
{\delta\hat U\over\delta x_\vk(\vJ)}\cos(\vk\cdot\vtheta)-
{\delta\hat U\over\delta y_\vk(\vJ)}\sin(\vk\cdot\vtheta)\Big)
{\p f_0\over\p\vJ}\cdot{\p\over\p\vtheta}\cr
&\sum_{\vk'}\Big(
{\delta\hat V\over\delta x_{\vk'}(\vJ)}\cos(\vk'\cdot\vtheta)
-{\delta\hat V\over\delta y_{\vk'}(\vJ)}\sin(\vk'\cdot\vtheta)\Big)
\cr
&={1\over2(2\pi)^3}\int\d^3\vJ
\sum_\vk\vk\cdot{\p f_0\over\p\vJ}\Big(
-{\delta\hat U\over\delta x_\vk}{\delta\hat V\over\delta y_\vk}+
{\delta\hat U\over\delta x_\vk}{\delta\hat V\over\delta y_\vk}\Big).\cr
\end{align}
To bring the Poisson
bracket \eqrf{eq:defsPB} to  the canonical form
\[\label{eq:canonPB}
\{\hat U,\hat V\}[f]=\int\d^6\vw\,\sum_\vk
\bigg({\delta\hat U\over\delta q_\vk}{\delta\hat V\over\delta p_\vk}
-{\delta\hat U\over\delta p_\vk}{\delta\hat V\over\delta q_\vk}\bigg)
\]
 we introduce new coordinates 
\[\label{eq:def_pq}
q_\vk\equiv u_\vk x_\vk\quad;\quad
p_\vk\equiv u_\vk y_\vk,
\]
where
\[
u_\vk\equiv \sqrt{{2{\rd(2\pi)^6}\over-\vk\cdot\p f_0/\p\vJ}}.
\]
and restrict the sum over $\vk$ to the half of $k$ space in which
$\vk\cdot\p f_0/\p\vJ<0$. 

To summarise: the Fourier coefficients $q_\vk$ and $p_\vk$ of DFs have
emerged as canonical coordinates for the space of DFs. If we assign a priori
probability to sets of DFs according to their volume elements
$\int\d^3\vJ\prod_\vk\d q_\vk\d p_\vk$, our assignment will be invariant
under canonical transformation to new coordinates for DFs, and, crucially,
the a priori probability of a set of DFs will remain unchanged as the DFs
evolve according to the CBE. We have shown that
$(\vq_\vk,\vp_\vk)$ are canonical coordinates only to linear order, but the
invariance of probabilities assigned by canonical coordinates is exact
because the full CBE is of Hamiltonian form.

\section{Ergodic systems}\label{sec:E}

Here we derive some results for stable ergodic systems -- equilibria with DFs
that are functions of the single-particle Hamiltonian $H(\vJ)$, so
$f_0(H(\vJ))$. {\rd\cite{Antonov1961} showed that to be stable the DF must
satisfy $f_0'(H)<0$.}

\subsection{Inner product}

For such systems the natural inner product of the space of DFs
is (e.g.\ LB21)
\[
\la f\ket g\equiv\int{\d^6\vw\over|f_0'(H)|}f^*(\vw)g(\vw).
\]
It is simple to show that expressed in terms of Fourier components the
product takes the form
\[\label{eq:prod_kk}
\la f\ket g=(2\pi)^3\int{\d^3\vJ\over|f_0'|}\sum_\vk f^*_\vk g_\vk.
\]

\subsection{Even and odd DFs}

\cite{Antonov1961} showed that it is useful to split the DFs of perturbed
ergodic systems into parts even and odd in $\vv$:
\[
f_\pm(\vx,\vv)\equiv\fracj12\big\{f(\vx,\vv)\pm f(\vx,-\vv)
\big\}.
\]
The unperturbed DF lies entirely within $f_+$ while $f_-$ is entirely due to
the perturbation. It is easy to show from the CBE that to first order in the
perturbation $f_\pm$ are related by
\[\label{eq:splitfdot}
\p_tf_+=-\vOmega\cdot\p_\vtheta f_-.
\]
In the case of a normal mode{\rd
\begin{align}
f(\vw,t)&=\sum_{\vk\cdot\vOmega>0}\Big(
f_\vk\e^{\i(\vk\cdot\vtheta-\omega t)}+f^*_\vk\e^{-\i(\vk\cdot\vtheta-\omega
t)}\Big),
\end{align}
where the sum is restricted to half of $\vk$ space to compensate for
the explicit inclusion of the complex conjugate of each term.
When we substitute this expansion in equation \eqrf{eq:splitfdot} and
equate coefficients of exponentials, we find
\[
\omega f_{\vk+}=\vk\cdot\vOmega f_{\vk-}\hbox{ and }
\omega f^*_{\vk+}=\vk\cdot\vOmega f^*_{\vk-}\ {(\vk\cdot\vOmega>0)}.
\]
Given that $f^*_{\vk}=f_{-\vk}$, a change of variable to $\vk'\equiv-\vk$ in
the  second relation yields
\[
\omega f_{\vk'+}=-\vk'\cdot\vOmega f_{\vk'-}\quad{(\vk'\cdot\vOmega<0)}
\] 
so we always have $\omega f_{\vk+}=|\vk\cdot\vOmega|f_{\vk-}$.} Hence,
\[\label{eq:minusTOtotal}
f_\vk=f_{\vk+}+f_{\vk-}=\Big({|\vk\cdot\vOmega|\over\omega}+1\Big)f_{\vk-}.
\]

\subsection{Energy of a mode}

The inner product has the dimensions of energy, and indeed LB21 show that the
energy of a van Kampen mode is
\[
E_{\rm mode}=\la f_-\ket{f_-}.
\]
Crucially, the energies of van Kampen modes are additive because their odd
DFs $f_-$ are mutually orthogonal.

Equations \eqrf{eq:prod_kk} and
\eqrf{eq:minusTOtotal} allow us to express $E$ in terms of the Fourier
components of the complete perturbed DF as
\[
E_{\rm mode}=(2\pi)^3\int{\d^3\vJ\over|f_0'|}\sum_\vk{|f_\vk|^2\over(1+|\vk\cdot\vOmega|/\omega)^2}.
\]
 Rewritten in terms of the canonical
variables \eqrf{eq:def_pq} the energy is
\[\label{eq:EmodeQP}
E_{\rm mode}
=\fracj12\int\d^3\vJ\sum_{\vk\cdot\vOmega>0}
{\vk\cdot\vOmega\over(1+|\vk\cdot\vOmega|/\omega)^2}\big(|q_\vk|^2+|p_\vk|^2
\big).
\]
$E_{\rm mode}$ has the form of the sum of the  Hamiltonians of harmonic
oscillators with frequencies $\vk\cdot\vOmega/(1+|\vk\cdot\vOmega|/\omega)^2$.
The energy of
the entire system, being a sum of the energies of individual modes, is also a
sum of harmonic-oscillator energies
\begin{align}
&E_{\rm total}=
\sum_{\rm modes}E_{\rm mode}\cr
&=\fracj12\int\d^3\vJ\sum_{\vk\cdot\vOmega>0}
\sum_{\rm modes}{\vk\cdot\vOmega\over(1+|\vk\cdot\vOmega|/\omega)^2}\big(
|q_\vk|^2+|p_\vk|^2\big).\cr
\end{align}

The key to computing thermodynamic potentials is to express the system's
Hamiltonian as a sum of harmonic-oscillator Hamiltonians because that done
the partition function (or the entropy $-\kB\ln\Omega$) is readily evaluated.
Equation \eqrf{eq:EmodeQP} brings us closer to that goal but unfortunately
not right to it: the energy of a single mode is given in terms of
contributions from many oscillators: each pair $(\vk,{\rd\vJ})$ corresponds to a
different oscillator.  LB21 give an expression for $E_{\rm mode}$ that
involves an integral over just resonant tori, i.e., ones satisfying
$\vk\cdot\vOmega=\omega$, but this integral involves the mode's potential
$\Phi[f]$, which is not easily computed. What's needed is a canonical
transformation from $(\vq_\vk[f],\vp_\vk[f])$ to new functionals
$(Q_i[f],P_i[f])$ such
that the energy of the $i$th mode is $\Omega_i(Q_i^2+P_i^2)$. Since the
Hamiltonian is a quadratic function in both systems, the sought-after
transformation could be linear. The new functionals $(Q_i[f],P_i[f])$ will
encode the amplitude and phase {\rd of the contribution from the $i$th van
Kampen mode}
that's required to build up the an arbitrary DF.

\cite{MorrisonShadwick} identified the required functionals in the case of a
one-dimensional, homogeneous, electrostatic plasma.  Unfortunately,
their treatment doesn't immediately generalise to multiple spatial dimensions
even in the case of a plasma.  Perhaps a similar transform could be found
that works in the three-dimensional case. If this could be done for an
electrostatic plasma, it could almost certainly be adapted to the very
similar case of the gravitating periodic cube \citep{BarnesGoodmanHut}, which
also offers scope for some interesting numerical experiments.

\section{Conclusions}\label{sec:conclude}

Galaxies and star clusters are in states that differ from the steady-state
solutions to the CBE that have traditionally been used to interpret data.
Differences between actual and idealised states can now be detected in
the most precise modern data, so we need to extend stellar dynamics so these
deviations are appropriately predicted. Predictions will generally be of a
statistical nature:  particular deviations will be assigned probabilities. 
The natural way to do this is to assign probabilities to individual DFs. This
needs to be done in a consistent manner. In particular, probabilities should
be not change when DFs evolve under the CBE. We have shown how this
requirement can be met by identifying a symplectic structure and associated
canonical coordinates for the space of DFs. 

The ideal canonical coordinates for DF space would be functionals that
determine the amplitude and phase of  each van Kampen mode that is required
to build up a given DF. These would be the angle-action cordinates of DF
space. We have not identified these functionals, but we have identified one
system of canonical coordinates, from which the angle-action coordinates might
be derived through a canonical transformation. We have shown, moreover, that
the energy of van Kampen mode is a quadratic function of the identified
coordinates. This amounts to a significant step on the road to a seductive
theory of the thermodynamics of stellar systems. 

\section*{Acknowledgements}

{\rd We thank Douglas Heggie for a careful and helpful referee's report.} Jun Yan Lau
gratefully acknowledges support from University College London's Overseas and
Graduate Research Scholarships.  James Binney is supported by the UK Science
and Technology Facilities Council under grant number ST/N000919/1 and by the
Leverhulme Trust through an Emeritus Fellowship. 

\section*{Data Availability}

No new data was generated or analysed in support of this research.

\bibliographystyle{mn2e} 
\bibliography{new_refs.bib}

\end{document}